\documentclass[pre,twocolumn,showpacs,superscriptaddress]{revtex4-2}
\usepackage{newtxtext,newtxmath}
\usepackage{microtype}
\usepackage{graphicx}
\usepackage{color}

\usepackage{url}
\usepackage[colorlinks=true,allcolors=blue]{hyperref}

\DeclareMathOperator{\Ei}{Ei}
\DeclareMathOperator{\erf}{erf}
\DeclareMathOperator{\erfc}{erfc}

\newcommand{\leibnizd}[1]{{\text d}{#1}}
\newcommand{\dr}{\leibnizd{r}}
\newcommand{\du}{\leibnizd{u}}
\newcommand{\dt}{\leibnizd{t}}

\newcommand{\Dr}{\Delta r}

\newcommand{\kB}{k_\text{B}}
\newcommand{\kT}{\kB T}

\newcommand{\lB}{l_{\text{B}}}
\newcommand{\kD}{\kappa_{\text{D}}}
\newcommand{\lD}{\lambda_{\text{D}}}

\newcommand{\ks}{\kD\sigma}

\newcommand{\phiref}{\phi^{\text{ref}}}
\newcommand{\phiass}{\phi^{\text{ass}}}
\newcommand{\gref}{g^{\text{ref}}}
\newcommand{\aNref}{a_N^{\text{ref}}}
\newcommand{\aNass}{a_N^{\text{ass}}}
\newcommand{\aNassid}{a_N^{\text{ass,id}}}
\newcommand{\aNassex}{a_N^{\text{ass,ex}}}

\newcommand{\half}{{\textstyle\frac{1}{2}}}
\newcommand{\quarter}{{\textstyle\frac{1}{4}}}

\newcommand{\naive}{na\"\i ve}
\newcommand{\role}{r\^ole}

\newcommand{\latin}[1]{{\itshape #1}}
\newcommand{\eg}{\latin{e.g.}}
\newcommand{\ie}{\latin{i.e.}}
\newcommand{\cf}{\latin{cf.}}
\newcommand{\etal}{\latin{et al.}}

\newcommand{\via}{\latin{via}}

\newcommand{\perse}{\latin{per se}}

\newcommand{\addendum}{\latin{addendum}}
\newcommand{\proviso}{\latin{proviso}}

\newcommand{\german}[1]{{\itshape #1}}
\newcommand{\ansatz}{\german{ansatz}}

\newcommand{\Eq}[1]{Eq.~\eqref{#1}}
\newcommand{\Eqs}[1]{Eqs.~\eqref{#1}}
\newcommand{\Fig}[1]{Fig.~\ref{#1}}
\newcommand{\Figs}[1]{Figs.~\ref{#1}}
\newcommand{\Refcite}[1]{Ref.~\onlinecite{#1}}

\newcommand{\Table}[1]{Table~\ref{#1}}

\newcommand{\Appendix}[1]{Appendix~\ref{#1}}

\newcommand{\partFig}[2]{Fig.~\hyperref[#1]{\ref*{#1}#2}}

\begin{document}

\title{Wertheim association theory for ion pairing in electrolytes: effect of neutral clusters}

\author{Patrick B. Warren}
\email{patrick.warren@stfc.ac.uk}
\affiliation{The Hartree Centre, UKRI Science and Technology Facilities Council, Daresbury Laboratory, Sci-Tech Daresbury, Warrington WA4 4AD, UK.}

\author{Andrew J. Masters}
\email{andrew.masters@manchester.ac.uk}
\affiliation{Department of Chemical Engineering, University of Manchester, Manchester M13 9PL, UK.}

\begin{abstract}
  We address the problem of the vapor-liquid phase transition in the restricted primitive model (RPM) using Wertheim’s statistical associating fluid theory to capture the effects of ion pairing which dominate the low-temperature vapor phase.  For this we employ a reference system in which ion-pairing is suppressed by a judicious modification of the interaction between unlike charges from $1/r$ to $\erf(\kappa r)/r$, where $\kappa$ is a state-dependent parameter chosen so that the Helmholtz free energy $A$ is at a null point ($\partial A/\partial\kappa=0$).  Unlike the original RPM, this reference fluid admits real solutions to the hypernetted-chain (HNC) closure of the Ornstein-Zernike equations over a wide range of densities and temperatures.  In the present study, we go beyond previous work [M.~Li, Ph.D.\ thesis, University of Manchester (2011)] to allow for isodesmic assembly of ion pairs into neutral clusters.  We find this has the potential to improve significantly the agreement with the Monte-Carlo results for the RPM vapor phase boundary.  We can also match recent results on anomalous underscreening in the RPM [H\"artel \etal, Phys.\ Rev.\ Lett.\ {\bf130}, 108202 (2023)] assuming that only the free ions contribute to the screening length.
\end{abstract}

\date{February 2018; April, May, October 2024; October 2025}

\maketitle

\section{Introduction}
The modern theory of electrolytes can perhaps be said to originate with Arrhenius' recognition in 1887 that in aqueous solution many salts dissociate into individual charged ions~\cite{arrhenius_1887}.  Later work by Debye and H\"uckel \cite{debye_1923}, and Bjerrum \cite{bjerrum_1926}, placed the theory of such so-called `strong' (\ie\ fully dissociated) electrolyte solutions on a firm foundation.  Recent interest in the field has been driven by pragmatic applications to improve the energy storage capacity of devices such as super-capacitors \cite{haertel_2017, kondrat_2023, lian_2020}, and in the development of novel devices in the emerging field of iontronics \cite{han_2022, *faez_2023, *bocquet_2023}.  Additionally, there has been renewed academic interest around the discovery of anomalous underscreening in electrolytes and ionic liquids in surface force balance experiments \cite{smith_2016, *lee_2017, *lee_2017b, *groves_2024, coupette_2018, adar_2019, coles_2020, zeman_2020, *zeman_2021, cats_2021, *kumar_2022, haertel_2023, jaeger_2023, *elliot_2024}

While the formulation of a good, equilibrium theory of electrolytes is always a challenge, the problems become particularly acute when, in contrast to aqueous solutions, the ions are in a melt or a low dielectric solvent.  In such cases the strong attraction between oppositely charged ions is significantly greater than the typical thermal energy and this leads to strong ionic association.  A well-known model which exhibits such behavior is the restricted primitive model (RPM), where the ions are represented as hard spheres with embedded point charges.  The point charges interact via Coloumb interactions, scaled by the relative permittivity of the background medium.  The RPM exhibits a low temperature  (\ie\ low background permittivity) vapor-liquid phase condensation transition \cite{orkoulas_1994, fisher_1994, stell_1995} in addition to sundry ordered phases at high packing fraction of lesser interest here.  Computer simulations indicate that while the liquid phase is a normal, disordered fluid, the vapor phase consists largely of ion pairs and quasi-neutral clusters \cite{caillol_1995, bresme_1995, luijten_2002, valeriani_2010, haertel_2023}.  Considerable theoretical and simulation effort has been devoted to the study of this phase transition, yet a fully satisfactory theory is still lacking.

The earliest theoretical approach to ion pair formation in strong electrolytes was due to Bjerrum who considered the system to consist of associated ion pairs in a quasi-chemical equilibrium with free ions \cite{bjerrum_1926}.  This basic idea has inspired many recent developments \cite{gillan_1983, fisher_1993, yeh_1996, guillot_1996, weiss_1998, adar_2017}. Still, the methodology is somewhat clumsy: the equilibrium does not arise naturally out of any overarching theory, and there is an arbitrariness in how an ion pair is defined.  There is also the question of how one incorporates the effects of higher order clusters that are observed in simulations \cite{caillol_1995, haertel_2023}.  To circumvent these problems, we extend an approach based on Wertheim association theory \cite{wertheim_1984a, *wertheim_1984b, li_2011} which removes both any arbitrariness in the definition of an ion pair, and which allows for systematic theoretical improvement \cite{*[{Our approach differs from RISM-like calculations based on solving Wertheim-Ornstein-Zernike equations; see }] [{}] jiang_2002}.  In addition it yields a physically transparent account of the vapor-liquid phase transition in the RPM that we extend to allow for higher-order cluster formation.

In what follows we shall first define the RPM, then briefly review the traditional quasi-chemical association models of ion pairing.  We will then introduce the Wertheim theory and show that it gives a good account of ion pairing at low densities.  We next show that with increasing density the theory predicts a vapor-liquid phase transition.  If we further allow for the formation of neutral clusters, we can construct a theory which matches the known vapor-liquid phase boundary.  We close with a discussion on the application of the theory to the phenomenon of anomalous underscreening.

\section{Restricted primitive model}
The restricted primitive model (RPM) comprises equal numbers $N/2$ of oppositely charged hard spheres (ions), with embedded point charges, immersed in a featureless dielectric continuum in a volume $V$, at a total density $\rho=N/V$.  The model is specified by the pair potentials
\begin{equation}
  \beta\phi_{ij}(r)=\Bigl\{\begin{array}{ll} \infty & (r\le\sigma)\,,\\
  \pm\,\lB/r & (r>\sigma)\,,
  \end{array}\label{eq:rpmpot}
\end{equation}
where the positive sign is taken if the ions carry the same charge, $\beta=1/\kT$ is the inverse of the unit of thermal energy, $\sigma$ is the hard sphere diameter, and $\lB = \beta q^2 / \epsilon$ is the Bjerrum length defined in terms of the charge $q$ on the ions and the medium dielectric permittivity $\epsilon$ (we work in cgs units).  The state space of the RPM is then completely specified by a reduced density and temperature,
\begin{equation}
  \rho^*=\rho\sigma^3\,,\quad
  T^*={\sigma}/{\lB}\,.\label{eq:rT^*}
\end{equation}
It will also prove helpful to define $\beta^*=\lB/\sigma=1/T^*$.

\begin{figure}
\begin{center}
\includegraphics{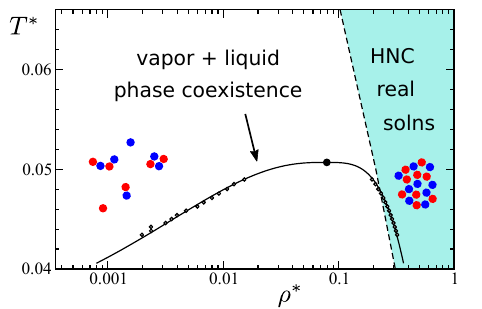}
\end{center}
\caption{Vapor-liquid phase coexistence in the restricted primitive model  (RPM) showing the coexistence region from Monte-Carlo simulations, redrawn from Luijten \etal\ \cite{*[{}] [{. The line in \Fig{fig:lims} is a fit to the Monte-Carlo data in this paper, assuming Ising universality and the law of rectilinear diameters in $\surd\rho^*$ (following up a suggestion by the authors). The fit expression is $\surd\rho^*=A+Bt\pm Ct^\beta$ where $t\equiv1-T^*/T^*_c\ge0$, the critical temperature $T^*_c=0.05069$, the exponent $\beta=0.326$, and the fit parameters are $A=0.2779$, $B=0.1925$, $C=0.4872$.}] luijten_2002}.  The solid marker is the critical point at $T^*\simeq0.0507$ and $\rho^*\simeq 0.079$.  The vapor (left) contains mostly ion pairs or quasi-neutral clusters \cite{caillol_1995, haertel_2023}, and coexists with a disordered liquid (right).  The region where the hyper-netted chain (HNC) closure to the Ornstein-Zernike (OZ) equations for the RPM has real solutions lies to the right of the dashed line \cite{lims-note}.}\label{fig:lims}
\end{figure}

Since water at room temperature has a Bjerrum length $\lB\simeq0.7\,\text{nm}$, and the ion sizes are also $O(1\,\mathrm{nm})$, aqueous 1:1 electrolytes typically correspond to the RPM at reduced temperatures $T^*\simeq0.5$--1.0 \cite{rasaiah_1972, attard_1993, haertel_2017}.  However as noted in the introduction there is much practical interest in multivalent electrolytes, electrolytes in low dielectric solvents, and ionic liquids.  In these systems it is often the case that the reduced temperature $T^*\alt0.1$.  At such temperatures ion pairing becomes significant, and the ion pairs and remaining unpaired ions may themselves assemble (transiently) into quasi-neutral clusters.  For $T^*\alt0.05$ the RPM exhibits the aforementioned vapor-liquid phase transition \cite{orkoulas_1994, fisher_1994, stell_1995}, where the vapor phase is comprised predominantly of these ion pairs and quasi-neutral clusters.  Despite decades of earlier work, accurate phase boundaries for this transition have only fairly recently been established by Monte-Carlo (MC) simulations~\cite{luijten_2002}.  The phenomenology is illustrated in \Fig{fig:lims}.

Let us first remark that since the RPM is completely specified by the pair potentials $\phi_{ij}(r)$ in \Eq{eq:rpmpot}, if we had an accurate liquid state theory for the corresponding pair functions $g_{ij}(r)$, we should have everything, including a full description of ion pair formation (without the need to introduce a quasi-chemical association equilibrium), and the vapor-liquid phase transition.  But, obviously, this just pushes the difficulties into the challenge of devising an accurate liquid state theory.  A `gold standard' in this area is perhaps the hypernetted chain (HNC) closure to the Ornstein-Zernike (OZ) equations \cite{hansen_2006}, which is known to be very accurate for the RPM at relatively high temperatures \cite{hansen_2006, vlachy_2011}, \cf\ \Fig{fig:bench}.  However, as is also well known \cite{belloni_1993}, real solutions to the HNC closure cease to exist at low temperatures and densities~\cite{*[{See \eg\ }][{}] lomba_1995}, as indicated in \Fig{fig:lims}.  Thus, whilst HNC can be used for the dense RPM liquid phase, the vapor phase is not easily accessible by this method.  

\section{Ion pairing}\label{sec:bjerrum}
In the RPM, the tendency to form ion pairs at low temperatures can be quantified by the contact energy between oppositely-charged ions $-\beta\phi_{+-}(\sigma)=\lB/\sigma=\beta^*$. For example, around the vapor-liquid critical point ($T^*\simeq0.05$) one has $\phi_{+-}(\sigma)\simeq-20\,\kT$, thus it is entirely unsurprising that the vapor phase contains mostly ion pairs or quasi-neutral clusters, with only small numbers of unassociated ions.  Obviously, a theory capable of describing the low temperature behavior of the RPM should take this into account.  However, ion pair formation is notoriously hard to handle systematically.  For example, in Bjerrum's original theory \cite{bjerrum_1926} an \ansatz\ has to be justified to define a quasi-chemical association constant $K$ which governs ion pair formation in this approach.  Furthermore, at the low temperatures of interest, we should also take into account deviations from ideality in the fluid of unpaired ions and dipoles.  For example if the Debye-Hückel (DH)~\cite{debye_1923} expressions for the ion activities are combined with Bjerrum's original \ansatz\ for $K$, one obtains the so-called DHBj theory of Fisher and Levin \cite{fisher_1993}.  A similar theory based on the ion activities from the exactly-soluble mean spherical approximation (MSA) was explored by Stell and coworkers~\cite{yeh_1996}.  

One important result from these quasi-chemical ion pairing theories is that as the density increases, the mole fraction $x$ of unpaired ions is expected to decrease.  This is a simple consequence of the Le Chatelier principle, reflecting the fact that increasing density (or equivalently pressure) shifts the quasi-chemical association equilibrium in the direction of reducing the number of unpaired ions.  However, $x$ may actually show a minimum (see \Fig{fig:xrho}), since increasing the density also depresses the ion activities. Such a minimum is seen for instance in DHBj theory around the critical point.  In addition, ion pairing may also be associated with physics not captured in these models.  For example, the ion pairs each have a dipole moment $\simeq q\sigma$, and at low temperatures it is at least plausible that the dipole-dipole attraction is strong enough to account for the additional clustering observed in these systems \cite{bresme_1995}, and perhaps even drive the condensation transition itself, noting that the phase behavior of dipolar hard dumbbells is very similar to the RPM \cite{shelley_1995}.  A related phenomenon arises from the collective polarisability of the ion pairs \cite{guillot_1996, weiss_1998, adar_2017}.  More formally, we can say that integrating out the dipolar degrees of freedom associated with the ion pairs should increase the background dielectric permittivity for the residual unpaired ions, \emph{diminishing} the propensity to form ion pairs.  On the other hand one can plausibly argue that only the long-range Coulomb forces are affected by this, so that ion pairing could be \emph{enhanced} because of the reduced non-ideality of the unpaired ions.  All these factors may be important and a systematic approach is clearly desirable. 

To set the scene, let us first discuss the phenomenological approach of Bjerrum and later workers, who assume a quasi-chemical association equilibrium between ion pairs and unpaired ions of opposite signs along the lines of
\begin{equation}
  \text{cation}\>+\>\text{anion}\>\rightleftharpoons\>\text{ion-pair}\,.
  \label{eq:qchem}
\end{equation}
Such models are predicated on the existence of a quasi-chemical association constant $K$ to describe the above equilibrium, and we shall use the Fisher-Levin DHBj model~\cite{fisher_1993} as an exemplar here.  For more details we refer to the careful study by Valeriani \etal\ \cite{valeriani_2010}.  To set the notation, we suppose that the monomer (unpaired ion) density is $\rho_1$ and the dimer (ion pair) density is $\rho_2$.  Since the system is overall neutral and ions are removed or added in pairs, the individual unassociated ion densities $\rho_1^+=\rho_1^-=\rho_1/2$.  We therefore have the constraints
\begin{equation}
  \rho=\rho_1+2\rho_2\,,\quad \rho_2
  =\quarter K\gamma_\pm^2\rho_1^2\,.\label{eq:qce}
\end{equation}
The first expresses the fact that the total ion density is conserved.  The second is the law of mass action for the quasi-chemical association equilibrium, including activity coefficients $\gamma_\pm$ for the unassociated ions.

Formally \cite{fisher_1993, valeriani_2010}, within this approach the association constant is the integral
\begin{equation}
  K=\int_\sigma^{R}\!4\pi r^2\dr \exp\Bigl(\frac{\beta^*\sigma}{r}\Bigr)\,.\label{eq:Keq}
\end{equation}
As is well documented, this integral diverges if the upper limit $R\to\infty$.  Hence, some \ansatz\ has to be introduced for the upper limit.  Bjerrum's original choice was $R=\lB/2$, which corresponds to the minimum of the integrand~\cite{bjerrum_1926}, and has been adopted by Fisher and Levin in the DHBj model~\cite{fisher_1993}.  It also fits with what Valeriani \etal~\cite{valeriani_2010} recommend.

We further follow Fisher and Levin \cite{fisher_1993} and change the integration variable in \Eq{eq:Keq} to $u=r/(\sigma\beta^*)$ to get
\begin{equation}
  K=\frac{4\pi\sigma^3Q\exp(\beta^*)}{\beta^*}\,,\quad Q
  =(\beta^*)^4\exp(-\beta^*)\!\int_{1/\beta^*}^{1/2}\!\!\du\,u^2e^{1/u}.\label{eq:Kdef}
\end{equation}
The latter is an $O(1)$ factor which can be written in terms of the exponential integral $\Ei(z)=\int_{-\infty}^z \dt\,e^{t}\!/t$,
\begin{equation}
\begin{split}
  Q={\textstyle\frac{1}{6}}
  [(\beta^*)^4\exp(-\beta^*)&(\Ei(\beta^*)-\Ei(2)+e^2)\\
  &\qquad{}-(\beta^*)^3-(\beta^*)^2-2\beta^*]\,.\label{eq:Qex}
  \end{split}
\end{equation}
For $\beta^*\agt5.69$ this is a decreasing function, from $Q\simeq3.36$ at the turning point and limiting to $Q\to1$ as $\beta^*\to\infty$.  

We now turn to the activity coefficients for the unassociated ions. For DH theory, introducing the Debye length $\kD^{-1}$ \via\ 
\begin{equation}
  \kD^2=4\pi\lB(\rho_1^++\rho_1^-)=4\pi\lB\rho_1\,,\label{eq:kD}
\end{equation}
and including a finite size effect, we have \cite{debye_1923, fisher_1993}
\begin{equation}
  \ln\gamma_\pm=-\frac{\beta^*\ks}{2(1+\ks)}\,.\label{eq:gDH}
\end{equation}
From \Eqs{eq:qce}, \eqref{eq:Kdef}, and \eqref{eq:gDH} therefore, one has
\begin{equation}
  \begin{array}{l}
    \displaystyle\rho=\rho_1+{\textstyle\frac{1}{2}}K\gamma_\pm^2\rho_1^2\,,
    \quad(\ks)^2={4\pi\rho_1\sigma^3\beta^*}\,,\\[6pt]
  \displaystyle K\gamma_\pm^2=\frac{4\pi\sigma^3 Q}{\beta^*}\,
    \exp\Bigl(\frac{\beta^*}{1+\ks}\Bigr)\,.
  \end{array}\label{eqs:DHBj}
\end{equation}
The exponential factor in the last term has been simplified by combining the first of \Eqs{eq:Kdef} with \Eq{eq:gDH}.

Taken together, \Eqs{eqs:DHBj} define the DHBj theory \cite{fisher_1993}.  For given values of $\beta^*=1/T^*$ and total density $\rho$, they should be solved self-consistently for the mole fraction of unpaired ions 
\begin{equation}
x=\rho_1/\rho\,.
\end{equation}  
As noted by Fisher and Levin \cite{fisher_1993}, the DHBj theory also contains a vapor-liquid phase transition with a critical point at $\ks=1$, where $T^*=1/16=0.0625$ and $\rho^*=(1/64\pi)(1+Q e^8/512)\simeq0.045$, where we used $Q\simeq1.39$ at $\beta^*=16$.  This is quite close to the real critical point, although the phase coexistence region itself has a peculiar `banana' shape \cite{fisher_1993}, which was attributed to the assumption of ideality for the ion pairs.  An alternative, which one might dub MSABj~\cite{yeh_1996}, uses the MSA activity coefficients \cite{*[{}] [{; with this theory \Eq{eq:gDH} is replaced $\ln\gamma_\pm = - \beta^*[1 + \ks - \sqrt{1+2\ks}] / \ks$.}] waisman_1970}.  However in terms of the prediction of ion pairing, for $\rho^*\alt0.1$ there is not much difference between these two theories.  

In the dilute limit the non-ideality of the unassociated ions may be neglected ($\gamma_\pm\to1$).  In this limit, \Eqs{eq:qce} reduce to
\begin{equation}
  1=x+\half K \rho x^2\,,\label{eq:prex}
\end{equation}
which is solved to obtain
\begin{equation}
  x=\frac{\sqrt{1+2K\rho}-1}{K\rho}\,.\label{eq:x}
\end{equation}
\Eqs{eq:Kdef}, \eqref{eq:Qex} and~\eqref{eq:x} provide simple expressions for the limiting behavior on dilution (see \Figs{fig:ktstar} and~\ref{fig:xrho}).

\section{Wertheim association theory}\label{sec:wertheim}
In a series of papers, Wertheim set up a formalism for dealing with association that avoided the use of arbitrary cut-offs in the definition of a cluster and proposed a thermodynamic perturbation theory (TPT) that permitted a rather straightforward way to calculate the thermodynamic properties of an associating fluid.  This approach, carried out for first order (TPT1) is extensively exploited for statistical associating fluid theory (SAFT) equations of state~\cite{chapman_1990, *mueller_2001, *economou_2002, *[{}][{ (this is the first in a series of such papers); }]avendano_2011, *lafitte_2013}. The model considered by Wertheim was that of a spherical particle with associating sites, or glue spots, on the sphere's surface. Steric effects would then prevent a higher degree of association than that given by the number of such sites.  Of course, the RPM has no such associating sites. That said, though, if we restrict ourselves, at least initially, to allow only for association into ion pairs (or dimers), we may still make use of the apparatus Wertheim set up.  We can then improve on this approximation by allowing the ion pairs to assemble isodesmically into neutral clusters.  We emphasise that as a matter of principle, the problem of defining an association constant encountered in the quasi-chemical association models simply does not arise in an approach using Wertheim's association theory.  This means that where the quasi-chemical models are phenomenological, our approach is systematic, and can be improved where deficiencies are identified.

\begin{figure}
\begin{center}
\includegraphics{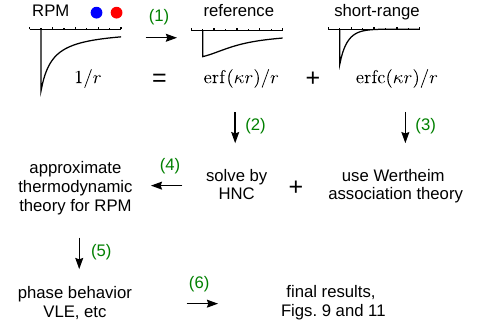}
\end{center}
\caption{Schematic of our strategy for solving the RPM: (1) the attraction between unlike charges is split into a weakened reference part plus a short-range correction; (2) the reference fluid is solved by HNC; (3) the short-range part is taken into account using Wertheim association theory; (4) the results are combined to make a thermodynamic theory for the RPM; (5) vapor-liquid phase coexistence boundaries are calculated; (6) the results are compared to the known phase diagram.\label{fig:diagram}}
\end{figure}

Our strategy (\Fig{fig:diagram}) is to split the attractive Coulomb interaction between oppositely-charged ions into a short-ranged and long-ranged contribution and treat the short-range part by Wertheim's theory, with the long-ranged part forming a reference system in which ion pair formation is suppressed.  A stationarity condition determines the splitting parameter.  Crucially, unlike the RPM, the HNC closure of the OZ equations for the reference fluid admits real solutions over a wide range of densities and temperatures.  With this approach therefore we can not only obtain predictions for the extent of ion pair formation, but we also have access to the full suite of the thermodynamic functions, which enable phase equilibria calculations. 

\subsection{Reference fluid}
\begin{figure}
\begin{center}
\includegraphics{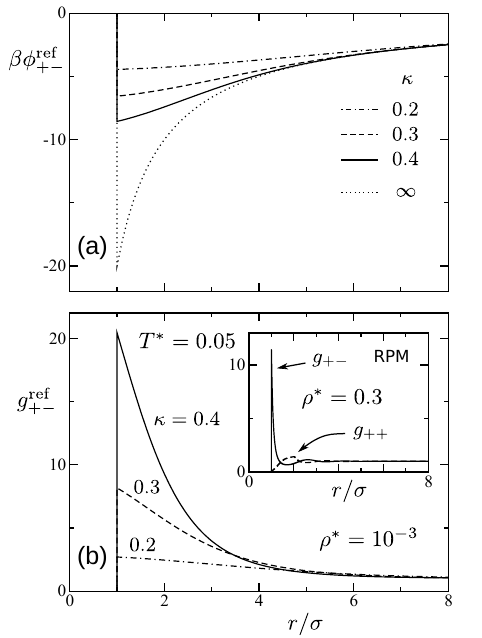}
\end{center}
\caption{Reference fluid: (a) potential between unlike ions for several values of $\kappa$ (the limit $\kappa\to\infty$ recovers the Coulomb law), and (b) the corresponding pair distribution functions at $T^*=0.05$ and $\rho^*=10^{-3}$ from HNC; for comparison the inset shows the HNC pair functions in the RPM liquid at the same reduced temperature and $\rho^*=0.3$.\label{fig:potpair}}
\end{figure}

The reference fluid is identical to the RPM specified in \Eq{eq:rpmpot} except that the attraction between oppositely charged ions is suppressed for $r\ge\sigma$ by writing the pair potential as
\begin{equation}
  \beta\phiref_{+-}=-\frac{\lB}{r}
  \erf\Bigl(\frac{\kappa r}{\sigma}\Bigr)
  \quad(r\ge\sigma)\,.\label{eq:refpot}
\end{equation}
In this $\kappa$ is a dimensionless splitting parameter (not to be confused with the inverse Debye length $\kD$), which we shall deal with in a moment.  Examples of this modified potential are shown in \partFig{fig:potpair}{a}, and the corresponding pair distribution functions, solved by HNC, are shown in \partFig{fig:potpair}{b}.  In the latter, the main plot shows $\gref_{+-}(r)$ in the reference fluid at $\rho^*=10^{-3}$.  Note the significant suppression of ion pair formation as $\kappa\to0$, signalled by the decline of $\gref_{+-}(\sigma)$.  This corresponds to the fact that the contact energy between unlike charges $-\beta\phi_{+-}(\sigma)=\beta^*\erf(\kappa)$ is reduced in magnitude compared to the RPM.  For example at $T^*=0.05$ and with $\kappa=0.2$, it is $\phi_{+-}(\sigma)\simeq -4.5\,\kT$.  The inset in \partFig{fig:potpair}{b} shows for comparison the HNC pair functions for the RPM in the liquid state at $\rho^*=0.3$\,; all calculations being at $T^*=0.05$.  More details of the HNC methodology are given in \Appendix{app:hnc}.

\subsection{Short range association contribution and ion pairing}
Having specified the reference fluid which accommodates the long range part of the potential split, we now turn to the treatment of the short range part.  For this we introduce
\begin{equation}
  \beta\phiass_{+-}=-\frac{\lB}{r}
  \erfc\Bigl(\frac{\kappa r}{\sigma}\Bigr)
  \quad(r\ge\sigma)\,,\label{eq:potass}
\end{equation}
such that the RPM is recovered as $\phi_{+-}=\phiref_{+-}+\phiass_{+-}$.
With this, we define the Wertheim integral
\begin{equation}
  \Delta_1=\int_0^\infty\!\!4\pi r^2\dr\, [\exp(-\beta\phiass_{+-})-1]\,
  \gref_{+-}(r)\,.\label{eq:dpm}
\end{equation}
Note $\gref_{+-}(r)=0$ for $r<\sigma$ in this so we do not need to specify $\phiass_{+-}$ within the hard cores.  In contrast to the quasi-chemical models, $\Delta_1$ is well defined.  This is a key advantage of the Wertheim approach.  Within the constraints of a one-site association model $\Delta_1$ plays the \role\ of $K\gamma_\pm^2$ in the previous section, \ie\ $\rho_2=\Delta_1\rho_1^+\rho_1^-$.  The mole fraction of unpaired ions then follows from
\begin{equation}
  1=x+\half\Delta_1\rho x^2\,.\label{eq:prewertx}
\end{equation}
This solves, \cf\ \Eqs{eq:prex} and~\eqref{eq:x}, to
\begin{equation}
  x=\frac{\sqrt{1+2\Delta_1\rho}-1}{\Delta_1\rho}\,.\label{eq:wertx}
\end{equation}
Equations~\eqref{eq:dpm}--\eqref{eq:wertx} are the essential results in this approach to ion pair formation, without including neutral clusters which we address below. We emphasise that they are rigorously based in Wertheim's thermodynamic perturbation theory.  The diagrammatic expansion of the dimer density has a series of terms, each containing an attractive $F$-bond between the dimer particles ($F = \exp(-\beta\phiref)[\exp(-\beta\phiass) - 1]$). Thus the definition itself does not involve a cut-off.  We rely on being able to solve for the pair function $\gref_{+-}$ in the reference fluid, and we need some method for fixing the splitting parameter $\kappa$ in order to be able to calculate $\Delta_1$.  This is discussed next.

\subsection{Splitting parameter}
In principle the final results should be insensitive to the splitting parameter $\kappa$ since changes in the reference potential should be compensated by changes in the Wertheim integral in \Eq{eq:dpm}.  To reflect this ideal, which is never perfectly attainable in an approximate theory, we fix $\kappa$ by imposing a stationarity condition on the free energy,
\begin{equation}
  {\partial a_N}/{\partial\kappa}=0\,.\label{eq:stat}
\end{equation}
Here $a_N=\aNref+\aNass$, the total free energy per ion, is the sum of the reference free energy which easily computed in HNC \cite{hiroike_1960}, and the contribution from the Wertheim association theory.  For a one-site association model \cite{wertheim_1984a, *wertheim_1984b}, the result can be obtained as a special case of the derivation sketched in \Appendix{app:assoc}.  The final result is, \cf\ \Eqs{eq:prena} and~\eqref{eq:na},
\begin{equation}
  \beta\aNass=\ln x + 1-x-\quarter{\Delta_1\rho x^2}
  = \ln x + \half(1-x)\,.\label{eq:werta}
\end{equation}

\begin{figure}
\begin{center}
\includegraphics{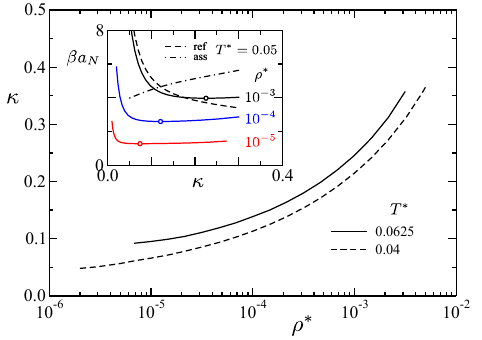}
\end{center}
\caption{Splitting parameter: dependence of $\kappa$ on $\rho^*$ according to \Eq{eq:stat} for two values of $T^*$ (main plot), and dependence of $\beta a_N$ on $\kappa$ for three values of $\rho^*$ at $T^*=0.05$ (inset; curves arbitrarily offset) with the stationary points (all minima) indicated by open circles.  Results are shown for the pure ion pairing model; the case where neutral clusters are included is similar.\label{fig:kappamin}}
\end{figure}

For a given state point, we calculate $\kappa$ following this prescription and this determines the Wertheim theory prediction for the extent of ion pairing and the thermodynamic functions at that point.  To illustrate this, we show in \Fig{fig:kappamin} the dependence of $\kappa$ on density at a couple of reduced temperatures, and in the inset the free energy as a function of $\kappa$ at a couple of representative state points.  We see that $\kappa\alt0.5$ typically, but also that it increases with increasing density.  This shifts the reference fluid back towards the RPM, where HNC fails to have real solutions.  As long as we demand that the stationarity condition in \Eq{eq:stat} should apply, this limits the applicability of our approach to low to moderate densities.  As the inset in \Fig{fig:kappamin} indicates, it turns out the stationarity condition corresponds to a minimum in the free energy, but we emphasise that there is no underpinning variational principle here.  Note also that the individual contributions $\aNref$ and $\aNass$ to the total free energy are decreasing and increasing functions of $\kappa$ respectively (\Fig{fig:kappamin} inset; dashed and chain lines), as the attractive potential well between unlike pairs shifts away from the short range part and back into the reference fluid.  Hence the stationarity condition corresponds to a balance between these.

\subsection{Inclusion of neutral clusters}\label{subsec:neut}
The question now arises of how best to include higher order clusters than dimers in our approach. The simplest procedure is to assume that clusters take the form of linear chains and that the free energy change of adding an ion to an existing cluster is independent of aggregation number. This is a simple isodesmic approximation. If we further assume that the equilibrium constant for all such additions is $\Delta_1$, then the resulting theory is equivalent to Wertheim's TPT1 but with each ion having two associating sites. We do not reproduce the result of such a calculation here, but suffice it to say that the results are far from encouraging!

One problem of this approach is that it is known from simulation that the majority of clusters are neutral, so we significantly overestimate the number of charged clusters, \ie\ we assume the existence of clusters with an odd number of ions which all have a net, overall charge. One may circumvent this problem by retaining the isodesmic approximation but only allowing neutral clusters to form.  Unfortunately, this approach (based on linear chains) also performs badly.

One big problem with this modified TPT1 theory is that simulation indicates that the clusters are not linear chains, but take up a variety of shapes. A second problem is that the magnitude of the free energy change on combining two charged ions is almost certainly greater than that of combining two neutral clusters. As discussed previously, in principle one may calculate the association constant for two neutral clusters using second order perturbation theory (TPT2), relaxing both the cluster shape and the isodesmic approximations noted above,  but it would be extremely challenging to carry out such a program rigorously. Such a methodology would require integrations over four particle distribution functions of the reference fluid. Not only is this numerically troublesome to apply in an iterative process, it also suffers from the problem of a lack of knowledge of the mathematical form of such correlation functions. Instead we now propose a more heuristic analysis which, we argue, provides helpful insights into the role of high order ion clusters.  For simplicity, but strongly motivated by the simulation results \cite{caillol_1995, haertel_2023}, we shall consider only the possibility of neutral clusters, formed from ion pairs.  As above we suppose the density of ion pairs is controlled by an association constant $\Delta_1$ but we now further allow these ion pairs to assemble into clusters, governed by a second (isodesmic) association constant $\Delta_2$. 

We therefore have
\begin{equation}
  \rho_2=\Delta_1\rho_1^+\rho_1^-\,,\quad
  \rho_{2n}=\Delta_2^{n-1}\rho_2^n\,,\label{eq:nclust}
\end{equation}
where $\rho_1^+$ and $\rho_1^-$ are the densities of the unassociated ions, and $\rho_{2n}$ ($n\ge1$) is the density of $(2n)$-mers, including $\rho_2$ for the remaining unassociated ion pairs.  The total number density of positive ions, for instance, is then
\begin{equation}
  \rho^+=\rho_1^++\sum_{n=1}^\infty\,n\,\rho_{2n}
  = \rho_1^+ +\frac{\rho_2}{(1-\Delta_2\rho_2)^2}\label{eq:nx1}
\end{equation}
(and similar for the negative ions).  We now inject the constraints $\rho_1^+=\rho_1^-=\rho_1/2$, where $\rho_1=x\rho$ is the number density of unpaired ions, $\rho^+=\rho^-=\rho/2$ for the densities of ions of each kind, and $\rho$ for the total number density of ions.  With this, \Eq{eq:nx1} becomes, \cf\ \Eq{eq:prewertx},
\begin{equation}
  1=x+\frac{\half\Delta_1\rho x^2}%
  {(1-\quarter\Delta_1\Delta_2\rho^2x^2)^2}\,.\label{eq:nx}
\end{equation}
Application of the Wertheim's methodology for TPT1 yields the association free energy density for this model.  We sketch out the details in \Appendix{app:assoc}, and the final expression is
\begin{equation}
    \beta\rho\aNass=
    \rho^+\ln\frac{\rho_1^+}{\rho^+}
    +\rho^-\ln\frac{\rho_1^-}{\rho^-}
    +\frac{\rho_2(1+\Delta_2\rho_2)}{(1-\Delta_2\rho_2)^2}\,.\label{eq:aNassclust}
\end{equation}
With the above definitions this reduces to
\begin{equation}
  \beta\aNass = \ln x+1-x-\frac{\Delta_1 \rho x^2}%
        {4(1-\quarter\Delta_1\Delta_2\rho^2x^2)}\,,\label{eq:prena}
\end{equation}
and with the aid of \Eq{eq:nx} further simplifies to, \cf\ \Eq{eq:werta},
\begin{equation}
  \beta\aNass = \ln x+\half(1-x)
  (1+\quarter\Delta_1\Delta_2\rho^2x^2)^{1/2}.\label{eq:na}
\end{equation}
\Eqs{eq:nx} and~\eqref{eq:na} comprise the theory here.  Setting $\Delta_2=0$ reduces to the previously considered case of neutral ion pair formation, in \Eqs{eq:prewertx} and~\eqref{eq:werta}.

\begin{figure}
\begin{center}
\includegraphics{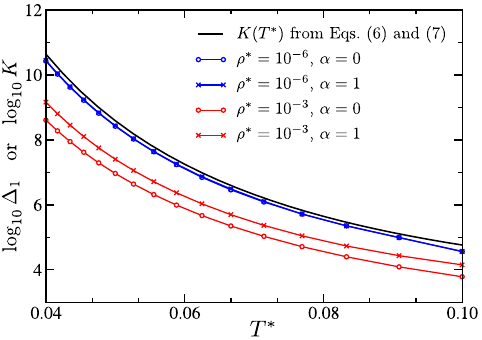}
\end{center}
\caption{Dependence of $\Delta_1$ on $T^*$ at two values of $\rho^*$ and two values of $\alpha$,  compared to the association constant $K(T^*)$ from the DHBj Bjerrum pairing model in \Eqs{eq:Kdef} and~\eqref{eq:Qex}.\label{fig:ktstar}}
\end{figure}
    
\subsection{Heuristic treatment of neutral cluster formation}
In order to introduce our approach to the assembly of neutral ion pairs, it is convenient to define a parameter $\alpha$, given by
\begin{equation}
  \Delta_2=\alpha^2\Delta_1\,,\label{eq:alpha}
\end{equation}
and examine the effect of varying $\alpha$.  We start by multiplying \Eq{eq:nx} through by $\Delta_1\rho$ and defining $z=\Delta_1\rho x$ to get
\begin{equation}
  \Delta_1\rho = z + \frac{z^2}{2(1-\quarter\alpha^2z^2)^2}\,.\label{eq:xy}
\end{equation}
This equation can be solved numerically for $z$ as a function of $\Delta_1\rho$, from which $x=z/(\Delta_1\rho)$ follows.  

We note that the right-hand side here behaves quite differently depending on whether $\alpha=0$ or $\alpha>0$: in the former, the expression always remains bounded; whereas in the latter, the expression blows up as $z\to2/\alpha$.  As a result, the solutions have different scaling behaviors for large $\Delta_1\rho$.  In the former case ($\alpha=0$), the solution is $z\sim({\Delta_1\rho})^{1/2}$ and consequentially $x\sim (\Delta_1\rho)^{-1/2}$ for $\Delta_1\rho\gg1$\,; \cf\ \Eq{eq:wertx}.  In the latter case ($\alpha>0$), the solution must remain bounded by $z<2/\alpha$, and as a consequence $x\sim(\Delta_1\rho)^{-1}$ for $\Delta_1\rho\gg1$.  This means that in the low temperature, low density limit, the density of unpaired ions scales quite differently with density in this model compared to the standard ion-pairing models.  

The association free energy per particle given in \Eq{eq:na} can be similarly written in terms of $x$ and $z$ as 
\begin{equation}
  \beta\aNass = \ln x+ \half(1-x)(1+\quarter\alpha^2z^2)^{1/2}.\label{eq:axy}
\end{equation}
To summarise, given the total ion density $\rho$ and $\Delta_1$, and for a given $\alpha$, we can solve \Eq{eq:xy} for $z$ and hence obtain $x$.  Substituting these into \Eq{eq:axy} gives the association free energy, which is combined with the reference fluid free energy computed as before, and extremised as a function of the splitting parameter.  The result is a model which incorporates ion pairing \emph{plus} neutral cluster assembly in a semi-systematic manner.

To sum up (see also \Fig{fig:diagram}), our starting point is to decompose the interaction between oppositely charged ions into a long-range and short-range interaction. The reference fluid, containing the long-range interaction is treated by HNC integral equation theory. In principle this may be improved via the use of superior closures. The short-range contributions are treated by Wertheim's thermodynamic perturbation theory. We carried this out to first order, but again, in principle, we could carry out the perturbation theory to higher order. We ignore non-neutral clusters, but again these effects may be systematically included, again within the TPT framework. Finally, related to the above, we assume isodesmic association for the neutral clusters, fitting the association constant to simulation data. Again, in principle, both these assumptions may be systematically improved.

\begin{figure}
\begin{center}
\includegraphics{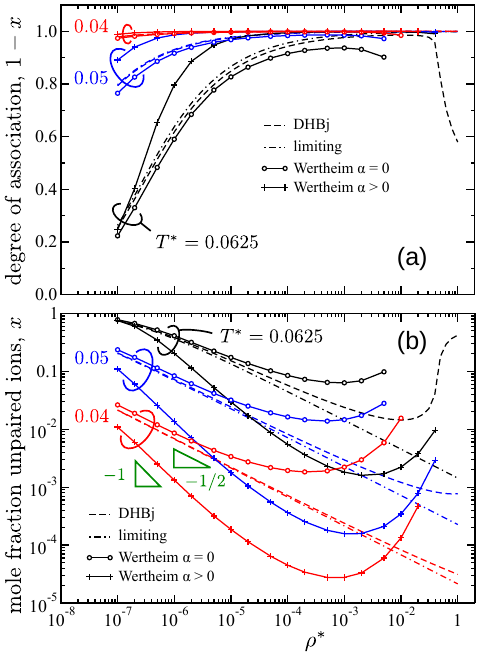}
\end{center}
\caption{Ion pairing in the RPM: (a) degree of association $1-x$ and (b) mole fraction $x$ of unpaired ions, plotted as a function of the overall density $\rho^*$, for three values of $T^*$ comparing $\alpha=0$ (no neutral clusters) with $\alpha>0$ (neutral clusters).  For the latter we chose $\alpha=0.05$ for $T^*=0.04$, $\alpha=0.5$ for $T^*=0.05$, and $\alpha=1$ for $T^*=0.0625$, to reflect the trend in the Arrhenius plot in \Fig{fig:arrhenius}.  Results are shown for the present Wertheim theory (solid lines), DHBj theory (dashed lines), and the limiting law behavior (chain lines).  Also indicated in (b) are the power laws $x\sim\rho^{-\nu}$ with $\nu=1/2$ and $\nu=1$.  \label{fig:xrho}}
\end{figure}

\section{Results}
\subsection{Ion pairing}
With this machinery in hand, we first investigate the predictions of the Wertheim theory for the extent of ion pairing, focusing on the low temperature vapor phase.  The first point to make is that for a large range of densities and temperatures, a value of the splitting parameter exists which meets the stationarity condition prescribed in \Eq{eq:stat}, so that the Wertheim theory does work.  Then, as shown in \Fig{fig:ktstar}, the Wertheim integral  $\Delta_1\sim10^5$--$10^{11}$ for the relevant state points, indicating a very strong tendency to form ion pairs.  Additionally $\Delta_1$ is a decreasing function of density, and of temperature.  The former can be attributed to the non-ideal behavior of the reference fluid, which captures the non-ideality of the unassociated ions.  The latter closely mimics the behavior of the Bjerrum pairing model.  Indeed, if we compare $\Delta_1$ with $K$ computed from \Eqs{eq:Kdef} and~\eqref{eq:Qex}, there is very good agreement at low density.  We also observe that allowing for neutral cluster formation ($\alpha>0$) lowers the ion pairing constant, but $\Delta_1$ is practically unaffected for $\rho^*=10^{-6}$.

\begin{figure}
\begin{center}
\includegraphics{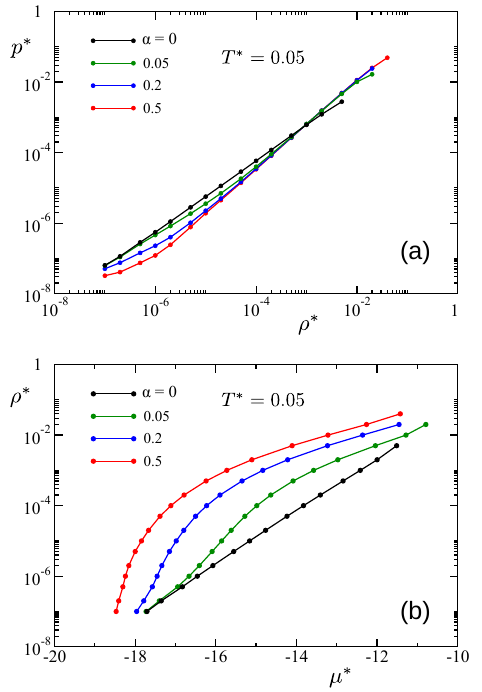}
\end{center}
\caption{Equation of state of Wertheim association model: (a) pressure $p^*=\beta p\sigma^3$ versus density, and (b) density versus chemical potential $\mu^* = \beta\mu$, for $T^*=0.05$ and the indicated values of $\alpha$.\label{fig:werteos}}
\end{figure}

The corresponding predictions for the extent of ion pairing are shown in \Fig{fig:xrho}, plotted both as the degree of dissociation $1-x$, and as the mole fraction $x$ of unpaired ions (on a double logarithmic scale).  As the temperature is reduced, $x$ diminishes, so that for $T^*=0.04$ and $\rho^*\sim10^{-3}$ for example, $\simeq99.8$\% of ions are in ion pairs.  However, unlike the limiting law, but seen in DHBj, there is a broad minimum in the mole fraction of unpaired ions around $\rho^*\sim10^{-4}$--$10^{-3}$.  In the DHBj theory this minimum occurs at higher densities and is a consequence of the non-ideality of the ionic system, which requires some model for the activity coefficients, but in the Wertheim theory it comes out naturally.  For the pure ion pairing case without neutral clusters ($\alpha=0$), our results are also systematically above the DHBj theory, which we attribute to the non-trivial capture of the ionic correlations beyond ion pair formation.  The effect of allowing for neutral clusters ($\alpha>0$) is to strongly depress the mole fraction of unpaired ions so that the prediction is now below the DHBj theory.  Moreover, the scaling in the regime where $\Delta_1\gg1$ changes from $x\sim\rho^{-1/2}$ to $x\sim\rho^{-1}$ (\partFig{fig:xrho}{b}), in accord with the discussion in section~\ref{subsec:neut}.  Allowing the possibility to form neutral clusters acts as a kind of `sink' for the ion pairs, qualitatively changing the scaling behavior.

\begin{figure}
\begin{center}
\includegraphics{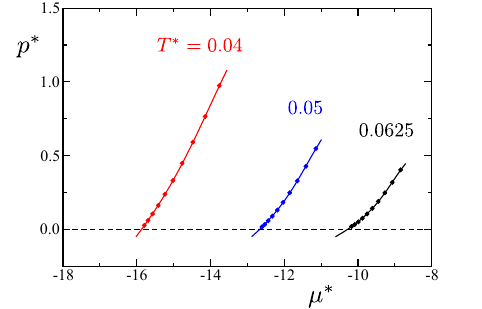}
\end{center}
\caption{Equation of state of RPM liquid according to HNC, showing the pressure as a function of the chemical potential, for three values of $T^*$. The points are the actual HNC results and the lines are a fit of $\mu^*(p^*)$ to a third-degree polynomial, extrapolated through $p^*=0$.\label{fig:rpmeos}}
\end{figure}

We believe our results compare relatively favourably to the careful study by Valeriani \etal\ \cite{valeriani_2010}, who looked at ion pair formation in the RPM using specialised Monte-Carlo methods which were able to access very low densities.  Their results adhere quite well to the quasi-chemical association model described in section~\ref{sec:bjerrum}, with Bjerrum's \ansatz\ for the association constant integral.  Our results (\Fig{fig:ktstar}) show that for $\rho^*=10^{-6}$, the Wertheim integral $\Delta_1$ is within 30--40\% of the quasi-chemical association constant $K$ evaluated using this theory, whether we take into account neutral clusters or not.  On the other hand with neutral cluster formation the predictions for the degree of dissociation (\partFig{fig:xrho}{a}) do show a deviation from the quasi-chemical association model (either DHBj or the limiting law), most prominantly for the $T*=0.0625$ curves.  However even these curves trends back onto the expected limiting law for density $\rho^*\alt10^{-6}$.  The qualitative change in the scaling behavior on allowing for neutral clusters becomes clear only for $x\alt10^{-2}$ (\partFig{fig:xrho}{b}).  The study by Valeriani \etal\ does not resolve this, since their technique is more focused on ion pair formation at very low densities rather than a very low value of the mole fraction of unpaired ions.

\subsection{Vapor-liquid phase coexistence}
We now turn our attention to the vapor-liquid phase transition in the RPM.  Here a key advantage of the Wertheim approach (with HNC for the reference fluid) is that the free energy $a_N=\aNref+\aNass$ is available.  To obtain the vapor pressure and chemical potential, we numerically differentiate this free energy with respect to overall density \cite{stat-note}.  As an example we show in \Fig{fig:werteos} the vapor pressure as a function of the density and the density as a function of the chemical potential.  The vapor pressure is approximately proportional to the density and is relatively insensitive to the inclusion of neutral clusters.  On the other hand incorporating neutral clusters markedly decreases the chemical potential.  This can be interpreted in the context of the reduction in the mole fraction of unpaired ions seen in \partFig{fig:xrho}{b}, since one might expect $\mu\simeq\ln\rho_1=\ln (x\rho)$, \ie\ set by the ideal chemical potential of the unpaired ions (this relation is exact, for an independent cluster model \cite{gillan_1983}).  

\begin{figure}
\begin{center}
\includegraphics{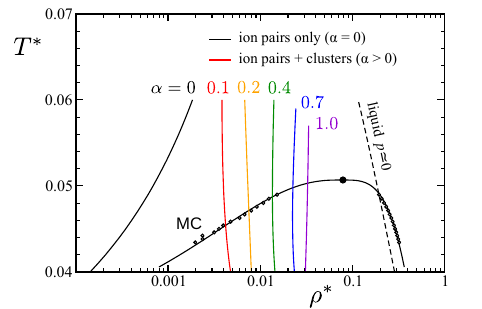}
\end{center}
\caption{Vapor phase boundaries from the Wertheim theory.  The liquid phase boundary (dashed line) is the locus where the RPM equation of state extrapolates to $p\simeq0$.  Monte-Carlo (MC) data as in \Fig{fig:lims}.\label{fig:coex}}
\end{figure}

\begin{figure}
\begin{center}
\includegraphics{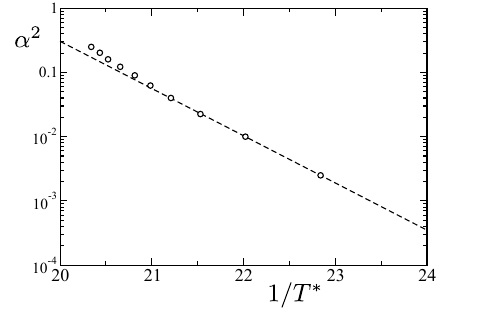}
\end{center}
\caption{Arrhenius plot of $\alpha^2$ as a function of $1/T^*$ where the Wertheim theory vapor phase boundary crosses the Monte-Carlo phase boundary in \Fig{fig:coex} (state points in \Table{tab:arrhenius}).\label{fig:arrhenius}}
\end{figure}

\begin{table}
  \centering
  \begin{tabular}{ccccccc}
    \hline
    $\alpha$ &\phantom{\quad}&
    $\rho^*$ &\phantom{\quad}&
    $T^*$ &\phantom{\quad}&
    $1/T^*$ \\
    \hline
    0.05 && 0.00247 && 0.0438 && 22.84 \\
    0.10 && 0.00421 && 0.0454 && 22.02 \\
    0.15 && 0.00593 && 0.0464 && 21.53 \\
    0.20 && 0.00753 && 0.0472 && 21.21 \\
    0.25 && 0.00899 && 0.0477 && 20.98 \\
    0.30 && 0.01038 && 0.0480 && 20.81 \\
    0.35 && 0.01196 && 0.0484 && 20.66 \\
    0.40 && 0.01357 && 0.0487 && 20.52 \\
    0.45 && 0.01486 && 0.0489 && 20.43 \\
    0.50 && 0.01643 && 0.0492 && 20.34 \\
    \hline
    \end{tabular}
  \caption{Intersection points $(\rho^*, T^*)$ of the Wertheim vapor phase boundary with the Monte-Carlo boundary, as a function of $\alpha$.\label{tab:arrhenius}}
\end{table}

The chemical potential is an increasing function of density (as must be the case for thermodynamic stability), and when it reaches the chemical potential of the dense RPM liquid, droplets of the liquid phase should condense out.  If we can estimate this point, we can predict the phase boundary on the vapor side of the vapor-liquid phase coexistence region.  To enable this we resort to the direct application of HNC (see \Appendix{app:hnc}) to the dense RPM liquid.  The limitations of our code restrict this to the region where real solutions can be found, as shown in \Fig{fig:lims}.   However, plotting the pressure as a function of chemical potential (\Fig{fig:rpmeos}) suggests we can apparently reliably extrapolate the chemical potential down to where $p\simeq0$, which we expect corresponds to the `boiling point' of the liquid, noting that the vapor pressure itself is very small (\partFig{fig:werteos}{a}).  Indeed, instead of doing a full coexistence calculation (which yields essentially the same results), we can use the $p\simeq0$ locus as an estimate of the liquid side of the vapor-liquid phase coexistence region.  We then extract the corresponding chemical potential (\Fig{fig:rpmeos}) and match it to the vapor phase chemical potential (\partFig{fig:werteos}{b}) to determine the vapor phase density.  In this way we can build up the phase boundaries predicted in this approach.

These results are shown in \Fig{fig:coex}.  In the absence of neutral clusters ($\alpha=0$), the predicted vapor phase boundary is too low.  Adding neutral clusters ($\alpha>0$) shifts the boundary to higher densities, where it starts to intersect the Monte-Carlo vapor phase boundary.  Stated another way, in the presence of neutral clusters, one needs to significantly increase the overall density to reach the same value of the chemical potential.  However we also note the vapor phase boundary is much steeper than the Monte-Carlo boundary.  The most obvious explanation is that our heuristic $\alpha^2=\Delta_2/\Delta_1$ is not a constant.

\begin{figure}
\begin{center}
\includegraphics{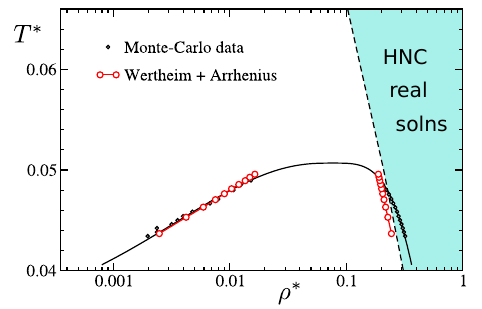}
\end{center}
\caption{Vapor-liquid phase boundaries from the Wertheim theory assuming the Arrhenius temperature dependence for $\alpha^2$ in~\Eq{eq:arr}.\label{fig:phasediag}}
\end{figure}

Physically, both association constants, $\Delta_1$ and $\Delta_2$, are proportional to $\exp(-\Delta G/\kT)$, where $\Delta G$ is the free energy change of associating a pair of ions and a pair of neutral clusters respectively. As noted earlier, these two free energies are hardly expected to have the same value. If we make the simplifying approximation that we can decompose $\Delta G$ into temperature independent enthalpies and entropies, one ends up with a predicted Arrhenius form for $\alpha^2$.  Indeed, one would in general expect this ratio to be temperature dependent if there is (as to be expected) a free energy difference between ion pair formation and assembly of the ion pairs into larger clusters.  To explore this we therefore adopted the following procedure.  For each value of $\alpha$, we determine the temperature $T^*$ where the Wertheim vapor phase boundary intersects the Monte-Carlo phase boundary.  This gives the sequence of state points $(\rho^*, T^*)$ shown \Table{tab:arrhenius}, as a function of $\alpha$.  If $\Delta_2/\Delta_1\propto\exp(-\Delta G/\kT)$ for some free energy difference $\Delta G$, this should be revealed in an Arrhenius plot of $\alpha^2$ against $1/T^*$.  This does indeed appear to be the case, as shown in \Fig{fig:arrhenius}.  The fit line therein is
\begin{equation}
  \ln\alpha^2=A-B/T^*\,,\label{eq:arr}
\end{equation}
where $A=32.7\pm0.3$ and $B=1.69\pm0.02$.  This result agrees with the notion that in the vicinity of the vapor-liquid phase transition the formation of a neutral cluster from ion pairs is energetically a little less favourable than the formation of the ion pairs themselves.  As the temperature is reduced, then $\alpha$ is also reduced, indicating a reduced tendency to form neutral clusters.  This however should be read in context of the overall increase in the tendency for ion pair formation as $T^*$ falls.

If we assume the Arrhenius temperature dependence of $\alpha$ as in \Eq{eq:arr}, and recompute the phase boundaries, the result is shown in \Fig{fig:phasediag}.  The agreement is gratifying as arguably there are only two adjustable parameters in the theory, namely $A$ and $B$ in the Arrhenius fit in~\Eq{eq:arr}.  However there have been many, perhaps judicious, choices made along the way, such as the assumption of isodesmic neutral cluster assembly from only ion pairs.  Nevertheless, the result can form the basis for a more refined approach.  

We note however that \Eq{eq:arr} predicts $\alpha>1$ for $T^*>B/A\simeq0.052$.  This is not precluded \perse\ in the theory, but becomes increasingly difficult to justify at high temperatures.  We also observe in \Fig{fig:arrhenius} that the data points are already starting to depart from the Arrhenius fit for $T^*\agt0.48$ (\ie\ $1/T^*\alt21$). Therefore we do not expect \Eq{eq:arr} necessarily to hold true for temperatures much above the vapor-liquid critical point at $T^*\simeq0.05$.  After all, enthalpy and entropy differences are not necessarily temperature independent, especially as the critical point is approached.

\subsection{Anomalous underscreening}
It has not escaped our attention that a theory such as ours, which predicts the extent to which `free' ions are lost to dimer and neutral cluster formation, can be leveraged to predict the screening behavior of the RPM as a function of density and temperature in the region where the theory has solutions.  Our theory therefore may be relevant to the debate about anomalous underscreening in electrolytes and ionic liquids mentioned in the introduction.  We first summarise the general situation.

Anomalous underscreening (AU) is usually taken to refer to the observation that aqueous electrolytes or ionic liquids usually show a monotonic exponentially-decaying tail in the force-distance curve in surface force balance experiments~\cite{smith_2016}.  At low ion densities or in weakly-coupled ionic liquids, the decay length associated to this matches the Debye length and fits with the expectations of DH theory.  At high salt concentrations or in strongly-coupled ionic liquids, the monotonic tail is often still present, typically after an initial oscillatory regime, but the decay length is anomalously large (AU regime) compared to what would be expected from DH theory.  
The experimental situation has not been unchallenged though, and until very recently it was unclear whether the RPM itself displayed any such analogous AU.  The situation has been clarified recently with the extensive simulations of H\"artel \etal\ \cite{haertel_2023} who find that the RPM can show AU in the pair distribution functions, albeit at a rather small relative amplitude (also found in experiment) and masked by simulation noise at high concentrations.

\begin{table}
  \centering
  \begin{tabular}{ccccccccc}
    \hline
     & & & && \multicolumn{4}{c}{mole fraction unpaired ions} \\[-3pt]
    $\text{conc}/\mathrm{M}$ & $10^3\rho^*$ & $\lD/\mathrm{nm}$ & $\lambda/\mathrm{nm}$
    && $(\lD/\lambda)^2$ && $\alpha=0$ & $\alpha=0.02$ \\
    \hline
    0.05 & 1.626 & 0.514 & 1.644(7) && 0.0978 && 0.0525 & 0.0410 \\
    0.07 & 2.276 & 0.434 & 1.973(3) && 0.0485 && 0.0564 & 0.0442 \\
    0.08 & 2.602 & 0.406 & 1.841(5) && 0.0487 && 0.0586 & 0.0461 \\
    0.09 & 2.927 & 0.383 & 1.816(2) && 0.0445 && 0.0610 & 0.0482 \\
    0.10 & 3.252 & 0.364 & 1.538(4) && 0.0559 && 0.0634 & 0.0505 \\
    \hline
    \end{tabular}
  \caption{Anomalous screening lengths from H\"artel \etal\ \cite{haertel_2023} at intermediate concentrations along the $T^*=0.06$ isotherm, with $\lB=5\,\mathrm{nm}$ and $\sigma=0.3\,\mathrm{nm}$.  The fifth column inteprets the anomalous screening length (fourth column) in terms of a mole fraction of unpaired ions.  The final two columns are from the present theory.  The Debye length $\lD$ (third column) is the `bare' value computed assuming all ions contribute, \ie\ $\rho_1=\rho$ in \Eq{eq:kD}.\label{tab:haertel}}
\end{table}

In terms of our theory, a simple though likely \naive\ approach is to assume the screening length $\lambda$ at a given state point is  given by the Debye length defined in \Eq{eq:kD} computed using the density $\rho_1$ of the \emph{free} ions.  Importantly, we note that this is an \addendum\ to the theory, which as it stands is otherwise self-contained, and unlike DHBj for example is not dependent on screening-related concepts.  The simulation results of H\"artel \etal\ afford an interesting test of this idea.  \Table{tab:haertel} shows the screening lengths extracted by H\"artel \etal\ at selected densities along the $T^*=0.06$ isotherm, where there is a clear signal of AU in the pair correlation functions~\cite{haertel-note}.  If we assume \Eq{eq:kD}, we can infer the mole fraction of unpaired ions from the measured screening length as $(\lD/\lambda)^2$, where $\lD$ is the `bare' Debye length computed assuming all ions contribute, \ie\ $\rho_1=\rho$.  The state points in \Table{tab:haertel} are accessible by our theory, and we compute the mole fraction of unpaired ions both in the absence of neutral clusters ($\alpha=0$), and in the presence of a small amount of neutral cluster formation ($\alpha=0.02$).  With the exception of the smallest concentration ($0.05\,\mathrm{M}$), our $\alpha=0$ theory underpredicts the inferred mole fraction of unpaired ions by around 20\%.  This is readily patched up by allowing for the indicated small amount of neutral cluster formation (but breaking significantly with \Eq{eq:arr} though).  Thus it is pleasing that in principle our theory can be matched to (a subset of) the recent simulation results for AU in the RPM.

Apart from this, note that the inferred mole fraction of unpaired ions (fifth column in \Table{tab:haertel}) is approximately constant or even increasing with density, in contradiction with the Le Chatelier principle (see discussion in section~\ref{sec:bjerrum}).  Such a retrograde trend is predicted to set in for large enough densities in our Wertheim theory (\partFig{fig:xrho}{b}), and also in DHBj albeit significantly later.  The outlier behavior of the measured screening length for the smallest concentration in \Table{tab:haertel} may then be a reflection of this crossover phenomenon.  We have limited ourselves to state points where AU is particularly clear in the simulation data though, and further analysis of the results in H\"artel \etal\ \cite{haertel_2023} might help clarify the trends here.

Regarding the experimental observations, the issue that both we and H\"artel \etal\ confront is that these are generally at higher densities ($\rho^*\simeq0.03$--0.26) and temperatures ($T^*\simeq0.07$--0.35) than either where AU in the RPM is directly measurable in simulation~\cite{haertel_2023}, or computable in our theory.  Such state points are in fact amenable to direct solution by HNC, but this unfortunately shows no sign of AU.  This suggests that the HNC, accurate as it appears to be for thermodynamic properties (see \eg\ \Fig{fig:bench}), is missing a subtle but critical aspect of the relevant underlying physics to account for AU.

\section{Discussion}
To summarise, we have applied Wertheim's association theory to ion pair formation in electrolytes in low dielectric solvents, using a novel potential splitting scheme in combination with a stationarity condition to fix the splitting parameter.  We apply this to the low temperature behavior of the RPM.  At low densities our result for the equivalent of the ion pairing association constant (\Fig{fig:ktstar}) is in good agreement with Bjerrum's theory as supported by the results of Valeriani \etal\ \cite{valeriani_2010}, and the densities of ion pairs (\Fig{fig:xrho}) are in reasonable agreement with existing models.  We used the theory to predict where a liquid phase would condense out of the vapor and found that agreement with the known phase boundary from Monte-Carlo simulations is improved (\Fig{fig:coex}) with the inclusion of neutral clusters of ion pairs.  This can be made quantitatively accurate (\Fig{fig:phasediag}) by the adoption of a suitable Arrhenius-like temperature dependence (\Fig{fig:arrhenius}) for the ratio of the ion pairing to neutral cluster formation association constants.  We were further able to match recent results on anomalous underscreening in the RPM \cite{haertel_2023} assuming that only the free ions contribute to the screening length, with the \proviso\ that the Arrhenius temperature dependence governing neutral cluster formation should be disapplied above the vapor-liquid critical point.

For our purposes, to account for cluster formation in the RPM we adopted a number of simplifying assumptions, namely that only neutral clusters are considered and that these assemble isodesmically from ion pairs.  Both these assumptions can be supported by old results from Caillol and Weis \cite{caillol_1995}, and indeed the more recent results from H\"artel \etal\ \cite{haertel_2023}.  These works find that indeed the vapour phase is dominated by ion pairs and mostly neutral clusters thereof.  Moreover the data from Caillol and Weis, although limited, can be interpreted as supporting an exponential size distribution of neutral clusters beyond ion pairs, indicative of isodesmic self-assembly; indeed they remark that the ``knowledge of only just two energy differences seems thus sufficient to predict the energy of a cluster of arbitrary size'' \cite{caillol_1995}.  This latter point tallies  with our finding that adopting an Arrhenius-like temperature dependence for the ratio of the association constants can fit the known phase boundary, indicating a constant free energy difference between the two stages of self-assembly.  We should emphasise that this result was in no way presupposed, rather it emerged from the theory without any preconceptions; we regard it as credible circumstantial evidence that our approach may be on the right track, given the heuristic exploratory elements.

Wertheim's theory has its roots in diagrammatic expansions and can be systematically improved.  The work presented here should be viewed as an preliminary step in this direction, particularly around the inclusion of neutral clusters.  The success of the approach and the discovery of the intriguing Arrhenius-like behaviour of the two association constants suggests that it is worth investing further effort to make a more detailed investigation along these lines.

\begin{acknowledgments}
We acknowledge many useful discussions over a long period with Fabian Coupette, Bob Evans, Michael Fisher, Denver Hall, Andreas H\"artel, George Jackson, Yan Levin, Susan Perkin, Ren\'e van Roij, and Josh Robinson, amongst others.  They have all helped develop and deepen our understanding of the rich behavior of the RPM and electrolytes in general.
\end{acknowledgments}

\appendix

\section{Numerical details of HNC}\label{app:hnc}
For the disordered liquid in the RPM, and the structure and thermodynamics of the reference fluid in the Wertheim association theory, we use a standard real-valued multi-component hyper-netted chain (HNC) closure of the Ornstein-Zernike (OZ) integral equations \cite{hiroike_1960}.  For this, we modified a HNC code developed originally for applications to soft potentials \cite{*[{}] [{. The HNC code itself is available as open source software at \url{https://github.com/patrickbwarren/SunlightHNC}.}] warren_2013}.  The code employs potential splitting methods to treat the long range electrostatics \cite{springer_1973, ng_1974}, and an accelerated convergence scheme originally proposed by Ng \cite{ng_1974, kelley_2004, vrbka_2009}.  We use standard literature expressions to compute the virial pressure, chemical potentials, and free energy \cite{hiroike_1960, lue_1994}.  

\begin{figure}
\begin{center}
\includegraphics{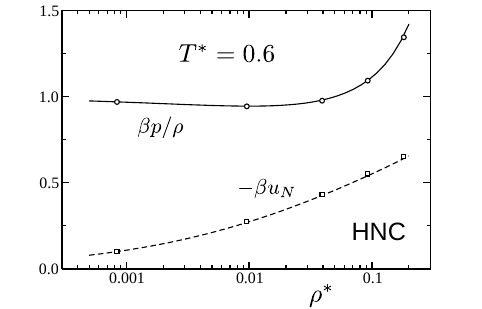}
\end{center}
\caption{HNC compressibility $\beta p/\rho$, and (negative) energy per particle $-u_N$ for the RPM at $T^*=0.6$ (lines), compared to Monte-Carlo results from Rasaiah \etal\ \cite{rasaiah_1972} (points); see also \Refcite{hansen_2006}.\label{fig:bench}}
\end{figure}

Several tests were made for numerical accuracy and to guard against coding errors.  First, the thermodynamic identity
\begin{equation}
  \rho a_N-{\textstyle\sum_i}\rho_i\mu_i+p=0
  \label{eq:test}
\end{equation}
should be verified exactly in HNC \cite{hiroike_1960}, and deviations reflect discretisation and truncation artefacts in the numerics.  We have checked this identity holds to a relative accuracy better than 1\% for all the calculations.  Second, we numerically differentiate the free energy to test the directly computed virial ressure and chemical potentials.  In all cases, errors are typically $\alt1$\%.  Finally, we have assured ourselves that we can recover known results for the HNC, such as those shown in \Fig{fig:bench}, or quoted in Table II in Bresme \etal\ \cite{bresme_1995}.

Our HNC code defines the pair functions on a uniform radial grid with grid spacing $\Dr$, $N_g$ grid points, and cut-off $R_g=N_g\,\Dr$.  For maximal efficiency of the fast Fourier transforms used in the code one would typically choose $N_g$ to be a power of 2.  For the reference fluid in the Wertheim theory we use $\Dr=10^{-3}\sigma$ and $N_g = 2^{16}=65\,536$ ($R_g\simeq 65$), but in order to get good results for the liquid state we have found that we need a very fine grid spacing to capture the steep variation in the pair functions at contact; in this case we use $\Dr=5\times10^{-5}\sigma$ and $N_g=2^{19}=524\,288$ ($R_g\simeq26$).

\section{Derivation of association free energy}\label{app:assoc}
Here we seek to provide a justification for \Eq{eq:aNassclust}, giving an expression for the association free energy system for our model, where oppositely charged monomers combine to form a neutral dimer, with an association constant $\Delta_1$, and the neutral dimers, in turn, aggregate isodesmically, with an association constant, $\Delta_2$, to generate higher order neutral clusters.  A strict derivation requires a modified version of Wertheim’s approach, but hopefully what follows here will give insights how the derivation works.

We split the free energy density change on association into ideal and excess parts,
\begin{widetext}
\begin{subequations}
\begin{align}
&\beta\rho\aNassid=
    \rho_1^+(\ln\Lambda^3\rho_1^+-1)
    +\rho_1^-(\ln\Lambda^3\rho_1^--1)+{\sum_{n=1}^\infty}\,\rho_{2n}(\ln\Lambda^3\rho_{2n}-1)
    -\rho^+(\ln\Lambda^3\rho^+-1)
    -\rho^-(\ln\Lambda^3\rho^--1)\,,\\
    &\beta\rho\aNassex=-{\sum_{n=1}^\infty}\,\rho_{2n}
    \Bigl[n\ln\frac{\Delta_1}{\Lambda^3}+(n-1)\ln\frac{\Delta_2}{\Lambda^3}\Bigr]\,.
\end{align}
\end{subequations}
The factors of $\Lambda^3$, where $\Lambda$ is the thermal de Broglie wavelength, ensures that the logarithms are all of dimensionless quantities but its value does not contribute to the final result.

The first equation corresponds to the ideal gas free energy change on converting a system of monomers into a mixture of monomers and clusters.  The second equation takes into account the bonding interactions between monomers once they form clusters. This bonding interaction corresponds to attractive $F$-bonds between monomers in a dimer and between dimers in a cluster. The first term in the square brackets corresponds to the interaction between the positive and negative monomers in each of the $n$-dimers that make up the cluster. The second term corresponds to the $(n-1)$ interactions between dimers as they assemble into neutral clusters.

The only awkward term present in these equations is the sum in the first equation. To cast this into a more convenient form, we note from \Eq{eq:nclust} that $\ln\rho_{2n} = (n-1)\ln\Delta_2 + n\ln\rho_2$.  Then to proceed, substitute $\rho_1^+=\rho^+-\sum_{n=1}^\infty n\,\rho_{2n}$ and likewise for $\rho_1^-$ into the last two terms in the first equation and gather terms involved in the sum over $n$.  Incorporating also the second equation yields
\begin{equation}
\begin{split}
&\beta\rho\aNass=
    \rho^+\ln\frac{\rho_1^+}{\rho^+}
    +\rho^-\ln\frac{\rho_1^-}{\rho^-}\\[3pt]
    &\qquad{}+\sum_{n=1}^\infty \rho_{2n}\Bigl[
    {}-n\,(\ln\Lambda^3\rho_1^+-1)
    -n\,(\ln\Lambda^3\rho_1^--1)
    +\ln\Lambda^3+(n-1)\ln\Delta_2+n\ln\rho_2-1
    -n\ln\frac{\Delta_1}{\Lambda^3}-(n-1)\ln\frac{\Delta_2}{\Lambda^3}
    \Bigr]\,.
\end{split}
\end{equation}
Substituting $\rho_2=\Delta_1\rho_1^+\rho_1^-$, the part of the summand in square brackets simplifies down dramatically to just $2n-1$.  Using \Eq{eq:nclust} again, it is then straightforward to evaluate the sum over $n$ analytically, and \Eq{eq:aNassclust} follows.
\end{widetext}

%\bibliography{rpm_selected}

\begin{thebibliography}{63}
\bibitem [{lim()}]{lims-note}%
  \BibitemOpen
  \href@noop {} {}\bibinfo {note} {Our own finding for the extent of the region
  where the HNC has real solutions agrees with what was reported by Belloni
  \cite{belloni_1993}. It is striking that the boundary of this region is just
  inside the locus where the HNC virial pressure vanishes.}\BibitemShut {Stop}%
\bibitem [{sta()}]{stat-note}%
  \BibitemOpen
  \href@noop {} {}\bibinfo {note} {We note the stationarity condition means we
  do not have to correct for the dependence of $\kappa$ on $\rho^*$ in taking
  these derivatives.}\BibitemShut {Stop}%
\end{thebibliography}

\end{document}